\documentclass[aps,prl,twocolumn,showpacs,superscriptaddress,groupedaddress]{revtex4}  
\pdfoutput=1
\usepackage{graphicx}  
\usepackage{amssymb}   
\usepackage{amsmath}

\usepackage{hyperref}
\usepackage{xcolor}

\def\gm{\gamma}
\def\bt0{\beta_{0}}
\def\bt1{\beta_{1}}
\def\bt2{\beta_{2}}
\def\bt3{\beta_{3}}
\def\gm0{\gamma_{0}}
\def\gm1{\gamma_{1}}
\def\gm2{\gamma_{2}}
\def\gm3{\gamma_{3}}
\def\nn{\nonumber}
\def\R{{\cal R}}
\def\l0{L_{\mu_0}}
\def\ol{\overline{\sigma}}

\begin{document}

\title{RG improved Higgs boson production to N$^3$LO in QCD}

\author{Taushif Ahmed}
\affiliation{The Institute of Mathematical Sciences, Chennai, India}

\author{Goutam Das}
\affiliation{Saha Institute of Nuclear Physics, Kolkata, India}

\author{M. C. Kumar}
\affiliation{The Institute of Mathematical Sciences, Chennai, India}

\author{Narayan Rana}
\affiliation{The Institute of Mathematical Sciences, Chennai, India}

\author{V. Ravindran}
\affiliation{The Institute of Mathematical Sciences, Chennai, India}

\date{\today}

\begin{abstract}
The recent result on the third order correction to the Higgs boson production through gluon fusion
by Anastasiou {\it et al.} \cite{Anastasiou:2015ema} not only provides a precise prediction with reduced scale uncertainties for
studying the Higgs boson properties but also establishes the reliability of the perturbative QCD.
In this letter, we propose a novel approach to further reduce the uncertainty arising from
the renormalization scale by systematically resumming the renormalization group (RG) accessible
logarithms to all orders in the strong coupling constant. Our numerical study based on this approach, demonstrates a
significant improvement over the fixed order predictions.
\end{abstract}

\maketitle
The remarkable discovery of the Higgs boson with a mass of about $125$ GeV by the ATLAS and CMS collaborations \cite{AtlasCMS} at the LHC has provided an important clue to understand the mechanism of spontaneous symmetry breaking within 
the framework of the Standard Model (SM) of particle physics. The technological advancements in experimental sectors augmented with the precise theoretical predictions, played crucial role in this distinctive discovery. But, with the new data to be available soon at the upgraded LHC, minimizing the theoretical uncertainties will be of paramount importance.
The pursuit of the precision studies in the Higgs boson production has been a consistent pioneer in advancing the perturbative QCD. 
It is worth recognizing the fact that the fixed order \cite{nnlo} as well as threshold resummed \cite{nnll} predictions in 
perturbative QCD along with the electroweak effects \cite{ewnnlo} played an important role not only in the exclusion of wide range of the Higgs boson masses but also to establish that the discovered boson is almost consistent with that of the SM.
%
%
Recent computation \cite{Anastasiou:2014vaa} of the complete threshold corrections at next-to-next-to-next-to leading order (N$^3$LO) including the $\delta (1-z)$ part has marked a milestone. Owing to the universality of the soft emissions, this result was followed by various new results \cite{oursn3loSV} for QCD processes at N$^3$LO in the threshold approximation. 
Very recently a state-of-the-art computation \cite{Anastasiou:2015ema} has been performed by Anastasiou {\it et al.} to accomplish
the complete N$^3$LO perturbative QCD correction to the inclusive Higgs boson production in the gluon fusion
channel.
This N$^3$LO corrected result not only demonstrates the reliability of the perturbation theory through the moderate correction, but also reduces uncertainties significantly resulting from renormalization ($\mu_R$) and factorization ($\mu_F$) scales in the range $\mu \in [\frac{m_{\rm H}}{4}, m_{\rm H}]$, where $m_{\rm H}$ is the mass of the Higgs boson. 
Up to next-to-next-to leading order (NNLO), it was demonstrated in \cite{Baglio:2010um} that there was a significant increase in scale uncertainties if we increase the range. We also observe a similar pattern even at N$^3$LO level for the $\mu_R$ variation.
This is because of the presence of large logarithms of the scale at every order.
Resumming such logarithms could often improve the scenario. 
In this letter, we use RG invariance of the Higgs boson production cross section to systematically 
resum these large logarithms to all orders in perturbation theory and show substantial reduction in the scale
uncertainties over the fixed order predictions.
In \cite{Ahrens:2008qu}, for the Higgs boson production, it was shown that the large corrections
of the form $(C_A \pi a_s)^2$ resulting
from analytical continuation of the form factors to time like
regions can be successfully resummed to all orders using RG, giving
rise to reliable predictions for $K$ factor.
Using effective field theory approach, the authors of \cite{Ahrens:2008nc} have
shown the role of RG in improving the theoretical predictions.
Our approach, while uses same RG invariance, differs from theirs
in treating the expansion parameter
in a systematic manner as it will be demonstrated in the following.

The inclusive hadronic cross section ($\sigma^{\rm H}(s, m_{\rm H}^2)$) for the Higgs boson production is related to the partonic cross-section ${\Delta}^{\rm H}_{a b} \left(\frac{\tau}{x_{1} x_{2}}, m_{\rm H}^2, \mu_{R}^{2}, \mu_F^2\right)$ as
\begin{align}
\label{Eq:CrossSec}
\sigma^{\rm H}(s, m_{\rm H}^2) =& \sigma^0 a_s^2 (\mu_R^2) \sum_{a, b} \int dx_{1} dx_{2} f_a (x_1,\mu_F^2) f_b (x_2,\mu_F^2)
\nonumber\\
\times& {\cal C}_{\rm H}^2\left(a_{s}(\mu_{R}^{2})\right) {\Delta}^{\rm H}_{a b} \left(\frac{\tau}{x_{1} x_{2}}, m_{\rm H}^2, \mu_{R}^{2}, \mu_F^2\right)
\end{align}
where, $f_a (x_1,\mu_F^2)$ and $f_b (x_2,\mu_F^2)$ are the parton distribution functions (PDFs), renormalized at $\mu_F$, of the initial state partons a and b with momentum fractions $x_1$ and $x_2$, respectively and  $\tau \equiv m_{\rm H}^2/s$ with $\sqrt{s}$ being the hadronic center of mass energy.
$a_s = \alpha_s/4 \pi$ with $\alpha_s$ being strong coupling constant, 
$\sigma^0$ is an overall factor describing the effective interaction between gluons and the Higgs boson at lowest order and ${\cal C}_{\rm H}$ is the Wilson coefficient.
Expressing 
$\sigma^{\rm H}(s, m_{\rm H}^2) = a_s^2 (\mu_R^2) ~ \overline{\sigma} (s, m_{\rm H}^2, \mu_R^2)$,
and using the RG invariance of $\sigma^{\rm H}(s, m_{\rm H}^2)$, namely $\mu_{R}^{2} \frac{d}{d \mu_{R}^{2}} \sigma^{\rm H}  = 0$,
we find
\begin{equation}
 \label{Eq:sigmaexp}
 \overline{\sigma} (\mu_{R}^{2}) = \overline{\sigma} (\mu_0^{2}) \exp \left[ - \int_{\mu_0^2}^{\mu_{R}^{2}} 
            \frac{d \mu^{2}}{\mu^{2}} \frac{2~\beta(a_s (\mu^2) )}{a_s (\mu^2)}  \right]
\end{equation}
where, 
$\beta(a_{s} (\mu^2) ) \equiv \mu^{2} \frac{d}{d \mu^{2}}a_{s}(\mu^{2}) = - \sum_{i=0}^{\infty} \beta_{i} ~a_{s}^{i+2} (\mu^2) \, .$
Considering $\mu_0$ as the central scale and using naive evolution of $a_s$, Eq.~\ref{Eq:sigmaexp} can be solved order by order
to obtain the following perturbative expansion of $\overline{\sigma} (\mu_{R}^{2})$
\begin{align}
 \label{Eq:sigmaexpansion}
 \overline{\sigma} (\mu_{R}^{2}) &= \sum_{n = 0}^{\infty} \sum_{k = 0}^{n} a_s^n (\mu_R^2) ~ {\cal R}_{n,k} ~ L_R^k
 \nn\\
           &= \sum_{n = 0}^{\infty} a_s^n (\mu_R^2) ~ \ol^{(n)} (\mu_{R}^{2})\,,
\end{align}
where $L_R = \ln \left( \frac{\mu_R^2}{\mu_0^2} \right)$. The coefficients of logarithms at each order in $a_s$,  
${\cal R}_{n,k} (0 < k \le n)$ are governed by the RG evolution and can be expressed in terms of the lower order ones,
${\cal R}_{n-1, 0}$, through
\begin{equation}
 \label{Eq:recursionrlns}
 \R_{n,n-m} = \frac{1}{(n-m)} \sum_{i=0}^{m} (n-i+1) \beta_i \R_{n-i-1,n-m-1}\,.
\end{equation}
The coefficient of the highest logarithms at $n^{th}$ order in $a_s$ grows 
as $(n+1) a_s^n \beta_0^n \R_{0,0}$
which often can give rise to potentially large contributions and can make the fixed order predictions unreliable. 
The RG invariance can be used to resum such contributions to all orders.
To achieve this task, we extend the approach of  
\cite{Ahmady:2002fd} to the case of scattering cross sections in hadron collisions. 
We rewrite Eq.~\ref{Eq:sigmaexpansion} as
\begin{align}
 \label{Eq:sigmresum}
 {\overline \sigma} (\mu_{R}^{2}) &= \sum_{m=0}^{\infty} a_s^m (\mu_R^2) ~ \sum_{n=m}^{\infty} {\cal R}_{n,n-m} ( a_s L_R )^{n-m}
\nn\\
                                  &= \sum_{m=0}^{\infty} a_s^m (\mu_R^2) ~ {\overline \sigma}_{\Sigma}^{(m)} \left( a_s (\mu_R^2) L_R \right)\,,
\end{align}
so that ${\overline \sigma}_{\Sigma}^{(m)}$ resums $a_s (\mu_R^2) L_R$ to all orders.
The closed form of ${\overline \sigma}_{\Sigma}^{(m)}$ can be obtained using RG invariance. 
The recursion relations (Eq.~\ref{Eq:recursionrlns})
which follow from the RG invariance,
can be used to show that ${\overline \sigma}_{\Sigma}^{(m)}$ satisfies the following first-order differential equations
\begin{align}
& \Big[ \omega \frac{d}{d \omega} + (m+2) \Big] \ol_{\Sigma}^{(m)} 
\nn\\
&= \Theta_{m-1} \sum_{i=1}^{m} \eta_i \Big[ (1-\omega) \frac{d }{d \omega} 
 - (m-i+2) \Big] \ol_{\Sigma}^{(m-i)}\,,
\end{align}
where $\Theta_{m-1}$ is Heaviside Theta function, $\eta_i \equiv \beta_i/\beta_0$ and $\omega = 1 - \beta_0 a_s (\mu_R^2) L_R$. 
Upon solving the above equations recursively, we obtain $\ol_m$ for all $m$. In Eq.~\ref{Eq:sigmaM}
we present them up to $m = 4$.

\begin{figure}[h!]
\centerline{
\includegraphics[width=9cm]{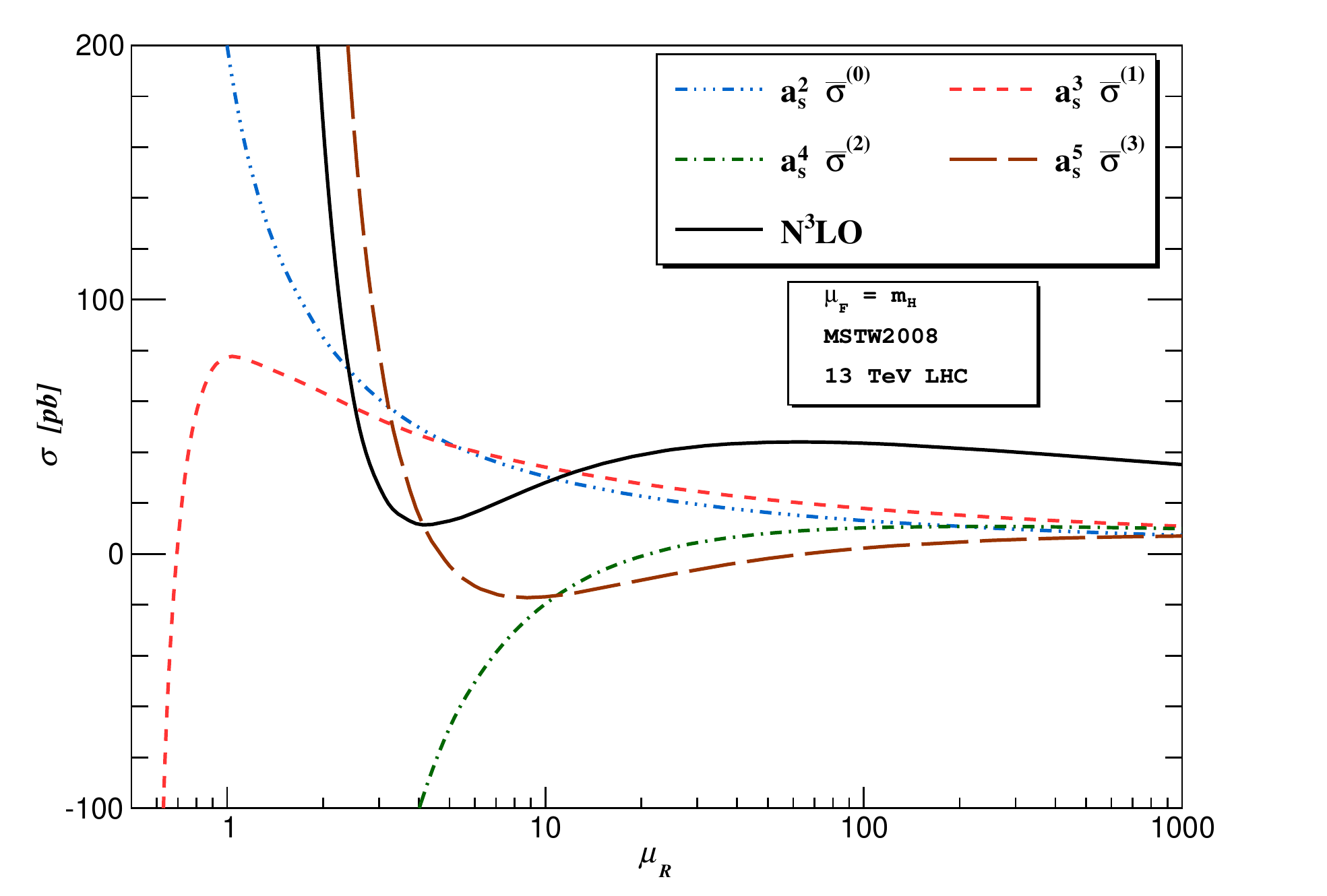}
}
\caption{\label{cbmur} 
${\mu_R}$ dependence of the LO and higher order corrections (FO) for LHC13, keeping $\mu_F = m_{\rm H}$ fixed.
}
\end{figure}

\allowdisplaybreaks
\begin{widetext}
\vspace{-0.5cm}
\begin{align}                        
%
%
&{\overline \sigma}_{\Sigma}^{(0)} 
= \frac{1}{\omega^2} \Big\{{\R_{0, 0}}\Big\}\, ,
%
%
~~ {\overline \sigma}_{\Sigma}^{(1)}
= \frac{1}{\omega^3} \Big\{ {\R_{1, 0}}-2 {\eta_1} {\R_{0, 0}} \ln(\omega) \Big\}\, ,
\nn\\ 
&{\overline \sigma}_{\Sigma}^{(2)} = \frac{1}{\omega^3} \Big\{ 2 {\R_{0, 0}} 
   \left( {\eta_1}^2 - {\eta_2}\right) \Big\}
+\frac{1}{\omega^4} \Big\{ {\R_{2, 0}} + 2 {\R_{0, 0}} \left( {\eta_2} - {\eta_1}^2\right)
+\ln(\omega) \Big(-2 {\eta_1}^2 {\R_{0, 0}}
-3 {\eta_1} {\R_{1, 0}}\Big)
+3 {\eta_1}^2 {\R_{0, 0}} \ln^2(\omega) \Big\}\, ,
\nn\\ 
&{\overline \sigma}_{\Sigma}^{(3)}
= \frac{1}{\omega^3} \Big\{{\R_{0, 0}} \left(
-{\eta_1}^3
+2 {\eta_1} {\eta_2}
-{\eta_3}\right) \Big\}
+\frac{1}{\omega^4} \Big\{ {\R_{0, 0}} \left(2{\eta_1}^3
-2 {\eta_1} {\eta_2}\right)
+{\R_{1, 0}} \left(3 {\eta_1}^2
-3 {\eta_2}\right)
+{\R_{0, 0}} \left(6 {\eta_1} {\eta_2}
-6 {\eta_1}^3\right) \ln(\omega) \Big\}
\nn\\
&~~~~\,
+\frac{1}{\omega^5} \Big\{{\R_{3, 0}} 
+ {\R_{0, 0}} \left({\eta_3}
-{\eta_1}^3\right) 
+{\R_{1, 0}} \left(3 {\eta_2}
-3 {\eta_1}^2\right) 
+\ln(\omega) 
\Big({\R_{0, 0}} \left(6 {\eta_1}^3
-8 {\eta_1} {\eta_2}\right)
-3 {\eta_1}^2 {\R_{1, 0}}
-4 {\eta_1} {\R_{2, 0}}\Big)
\nn\\
&~~~~\,
+\ln^2(\omega) \Big(7 
{\eta_1}^3 {\R_{0, 0}}
+6 {\eta_1}^2 {\R_{1, 0}}\Big)  
- 4 {\eta_1}^3 {\R_{0, 0}} \ln^3(\omega) \Big\}\,,
\nn\\
&{\overline \sigma}_{\Sigma}^{(4)}
= \frac{1}{\omega^3} \Big\{ \R_{0, 0} \Big( \frac{2}{3} \eta_1^4
- 2 \eta_1^2 \eta_2
+ \frac{2}{3} \eta_2^2
+ \frac{4}{3} \eta_1 \eta_3
- \frac{2}{3} \eta_4 \Big) \Big\}
+  \frac{1}{\omega^4} \Big\{ \R_{0, 0} \Big( 2 \eta_1^4
- 4 \eta_1^2 \eta_2
+ 3 \eta_2^2
- \eta_1 \eta_3 \Big)
\nn\\
&~~~~\,
+
\R_{0, 0} \ln(\omega)     \Big( 3 \eta_1^4
- 6 \eta_1^2 \eta_2
+ 3 \eta_1 \eta_3 \Big)
+
\R_{1, 0}  \Big( -\frac{3}{2} \eta_1^3
+ 3 \eta_1 \eta_2
- \frac{3}{2} \eta_3 \Big)  \Big\}
+ \frac{1}{\omega^5}
  \Big\{ \R_{0, 0} \Big(
- 6 \eta_1^4
+ 14 \eta_1^2 \eta_2
- 8 \eta_2^2 \Big)
\nn\\
&~~~~\,
+ \R_{0, 0} \ln^{2}(\omega)  \Big( 12 \eta_1^4
- 12 \eta_1^2 \eta_2 \Big)
+ \R_{1, 0} \Big( 3 \eta_1^3
- 3 \eta_1 \eta_2 \Big)
+
    \ln(\omega) \Big[ \R_{0, 0} \Big(
- 14 \eta_1^4
+ 14 \eta_1^2 \eta_2 \Big)
+ \R_{1, 0} \Big(
- 12 \eta_1^3
+ 12 \eta_1 \eta_2 \Big)
        \Big]
\nn\\
&~~~~\,
+ \R_{2, 0} \Big( 4 \eta_1^2
- 4 \eta_2 \Big)  \Big\} 
+
  \frac{1}{\omega^6} \Big\{ \R_{0, 0} \Big( \frac{10}{3} \eta_1^4
- 8 \eta_1^2 \eta_2
+ \frac{13}{3} \eta_2^2
- \frac{1}{3} \eta_1 \eta_3
+
      \frac{2}{3} \eta_4 \Big)
+ 5 \eta_1^4 \R_{0, 0} \ln^{4}(\omega) 
+    \R_{1, 0} \Big( -\frac{3}{2} \eta_1^3
+ \frac{3}{2} \eta_3 \Big)
\nn\\
&~~~~\,
+ \ln^{3}(\omega) \Big( -\frac{47}{3} \eta_1^4 \R_{0, 0}
- 10 \eta_1^3 \R_{1, 0} \Big)
+
\R_{2, 0}    \Big(
- 4 \eta_1^2
+ 4 \eta_2 \Big) 
+ \ln^{2}(\omega) \Big[ \R_{0, 0} \Big(
- 8 \eta_1^4
+ 20 \eta_1^2 \eta_2 \Big) 
+
      \frac{27}{2} \eta_1^3 \R_{1, 0}
\nn\\ 
&~~~~\,
+ 10 \eta_1^2 \R_{2, 0} \Big]
+
    \ln(\omega) \Big[ \R_{0, 0} \Big( 11 \eta_1^4
- 8 \eta_1^2 \eta_2
- 5 \eta_1 \eta_3 \Big)
+
\R_{1, 0}      \Big( 12 \eta_1^3
- 15 \eta_1 \eta_2 \Big) 
- 4 \eta_1^2 \R_{2, 0}
- 5 \eta_1 \R_{3, 0} \Big]
+ \R_{4, 0} \Big\} \,.
\label{Eq:sigmaM}
%
\intertext{
Alternatively $\ol_{\Sigma}^{(m)}$ can be computed from Eq.~\ref{Eq:sigmaexp} using RG improved solution for $a_s$, given in Eq.~\ref{Eq:asresum},
which implicitly resums the large logarithmic contributions to all orders in the perturbation theory.
}
%
%
& a_s (\mu_0^2) = a_s (\mu_R^2) \frac{1}{\omega} 
              + a_s^2 (\mu_R^2) \Bigg[ \frac{1}{\omega^2} \left( - \eta_1 \ln \omega \right) \Bigg]
              + a_s^3 (\mu_R^2) \Bigg[ \frac{1}{\omega^2} \left( \eta_1^2 - \eta_2 \right)
                                + \frac{1}{\omega^3} \left( - \eta_1^2 + \eta_2 - \eta_1^2 \ln \omega
                                                           + \eta_1^2 \ln^2 \omega \right)  \Bigg] 
\nn\\
& \hspace{1cm}
            + a_s^4 (\mu_R^2) \Bigg[ 
             \frac{1}{\omega^2} \Big\{ - \frac{1}{2} \eta_1^3 + \eta_1 \eta_2 - \frac{1}{2} \eta_3 \Big\} 
            +\frac{1}{\omega^3} \Big\{ \eta_1^3 - \eta_1 \eta_2 + \Big( - 2 \eta_1^3 + 2 \eta_1 \eta_2 \Big) \ln(\omega) \Big\}  
            +\frac{1}{\omega^4} \Big\{ - \frac{1}{2} \eta_1^3 + \frac{1}{2} \eta_3 
\nn\\  
& \hspace{1cm}
            + \Big( 2 \eta_1^3 - 3 \eta_1 \eta_2 \Big) \ln(\omega) + \frac{5}{2} \eta_1^3 \ln^{2}(\omega) - \eta_1^3 \ln^{3}(\omega) \Big\}
                                \Bigg]
             + a_s^5 (\mu_R^2) \Bigg[
               \frac{1}{\omega^2} \Big\{ \frac{1}{3} \eta_1^4 - \eta_1^2 \eta_2 + \frac{1}{3} \eta_2^2 + \frac{2}{3} \eta_1 \eta_3 - \frac{1}{3} \eta_4 \Big\}  
\nn\\   
& \hspace{1cm}
              + \frac{1}{\omega^3} \Big\{ \frac{1}{2} \eta_1^4 - \eta_1^2 \eta_2 + \eta_2^2 - \frac{1}{2} \eta_1 \eta_3 +
                       \Big( \eta_1^4 - 2 \eta_1^2 \eta_2 + \eta_1 \eta_3 \Big) \ln(\omega) \Big\} 
              + \frac{1}{\omega^4} \Big\{ - 2 \eta_1^4 + 5 \eta_1^2 \eta_2 - 3 \eta_2^2 
\nn\\      
& \hspace{1cm}
              + \Big( - 5 \eta_1^4 + 5 \eta_1^2 \eta_2 \Big) \ln(\omega) 
              + \Big( 3 \eta_1^4 - 3 \eta_1^2 \eta_2 \Big) \ln^{2}(\omega) \Big\} 
             +\frac{1}{\omega^5} \Big\{ \frac{7}{6} \eta_1^4 - 3 \eta_1^2 \eta_2 + \frac{5}{3} \eta_2^2 - \frac{1}{6} \eta_1 \eta_3 + \frac{1}{3} \eta_4 
\nn\\
\label{Eq:asresum} 
& \hspace{1cm}
             + \Big( 4 \eta_1^4 - 3 \eta_1^2 \eta_2 - 2 \eta_1 \eta_3 \Big) \ln(\omega) +
    \Big( -\frac{3}{2} \eta_1^4 + 6 \eta_1^2 \eta_2 \Big) \ln^{2}(\omega) - \frac{13}{3} \eta_1^4 \ln^{3}(\omega) +
    \eta_1^4 \ln^{4}(\omega) \Big\}
                               \Bigg]\,.  
\end{align}
 \end{widetext}
In the above Eq.(\ref{Eq:asresum}), the terms up to $a_s^4$ is already known, \cite{Moch:2005ba} and $a_s^5$ term is obtained
for the first time.
%
%
In the following, we study the numerical impact of fixed order (FO) as well as RG improved resummed (RESUM) 
cross sections up to N$^3$LO in QCD for the Higgs boson production through gluon fusion at the LHC.  
We have used an in-house Fortran code to do this.  We set $\mu_0=\mu_F=m_{\rm H}=125$ GeV throughout and use 
MSTW2008nnlo \cite{Martin:2009iq} parton distribution functions with the corresponding strong coupling constant 
from LHAPDF \cite{Whalley:2005nh}, $\alpha_s(m_Z) = 0.11707$.
At LO, the exact top and bottom quark mass effects are included through $\sigma^0$ in Eq.~\ref{Eq:CrossSec}.
Finite quark mass effects at NLO are taken into account using {\tt iHixs} at $\mu_R = \mu_F$.
At NNLO and N$^3$LO, we use effective theory predictions in the large top quark mass limit.
We first obtain $L_R$ independent terms namely $R_{0,0}, R_{1,0}$ and $R_{2,0}$ 
by setting $\mu_R = m_{\rm H}$ in our code whereas $R_{3,0}$ is extracted from the recent result 
for N$^3$LO cross section given in \cite{Anastasiou:2015ema} for the same choice of $\mu_R = \mu_F = m_{\rm H}$.
These $R_{i,0},~ (i=0,1,2,3)$ thus obtained at $\mu_F=m_{\rm H}$ with MSTW2008nnlo are
the only required ingredients to study the $\mu_R$ dependence of both the FO (Eq.~\ref{Eq:sigmaexpansion}) and the 
RESUM (Eq.~\ref{Eq:sigmresum}) cross sections up to N$^3$LO in QCD.
Note that the coefficients of all the $L_R$'s in Eq.~\ref{Eq:sigmaexpansion} can be obtained using
the recursion relations (Eq.~\ref{Eq:recursionrlns}).
As it was demonstrated in \cite{Anastasiou:2015ema}, 
inclusion of N$^3$LO corrections makes 
the $\mu_R$ sensitivity of the cross section milder compared to NNLO corrected results
when the $\mu_R$ is
taken to be closer to $m_{\rm H}$, say between $m_{\rm H}/4$ and $2 m_{\rm H}$. On other hand, if we
decrease $\mu_R$ below $m_{\rm H}/4$, the contributions from $L_R$ increase substantially
surpassing the scale independent ones giving rise to potentially large
scale uncertainties.  This happens at every order in perturbation
theory and the renormalization scale at which this happens, increases with the order. 
In Fig.~\ref{cbmur}, we quantify this up to N$^3$LO for FO.

\begin{figure}[h!]
\centerline{
\includegraphics[width=0.5\textwidth]{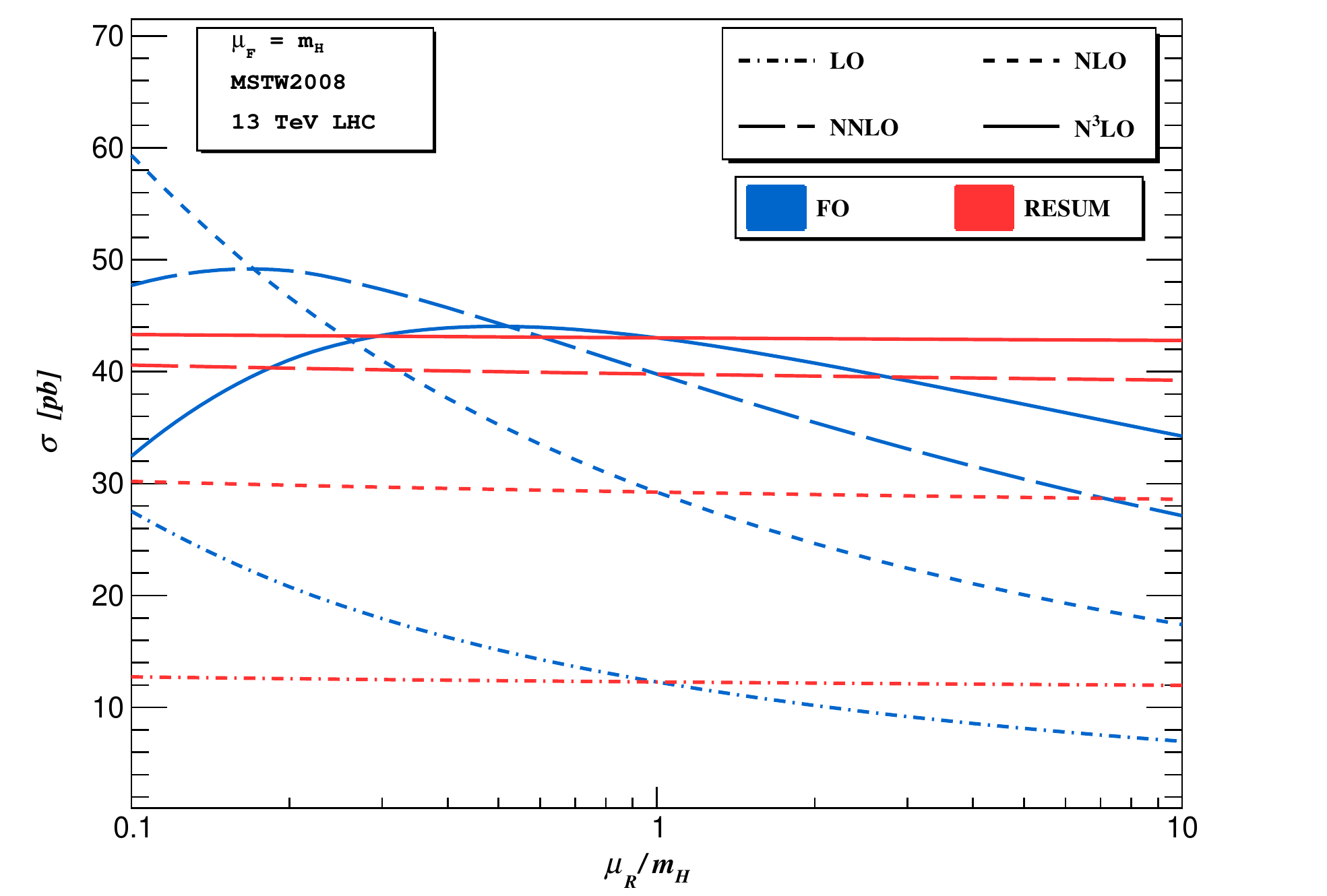}
}
\caption{\label{mur-var} 
${\mu_R}$ dependence of both the fixed order and resummed cross sections 
up to N$^3$LO.
}
\end{figure}

In the FO results (Eq.~\ref{Eq:sigmaexpansion}), the dependence on $\mu_R$ enters through the 
evolution of $a_s (\mu_R^2)$ as well as the perturbative
corrections that are polynomials in $L_R$ of the order $k \le n$ consistent with RG invariance. 
As $\mu_R$ decreases, the coupling constant as well as the magnitude of $L_R$ will increase, consequently,
for $\mu_R$ much less than $m_{\rm H}$, the contributions of the kind $a_s^n \beta_0^k L_R^k$ can 
become large enough to make the $\mu_R$ dependent terms even negative. 
Moreover, at higher orders, the contribution
of polynomial in $L_R$ need not be monotonic, 
instead it can change its sign with decreasing $\mu_R$ as seen in fig.\ref{cbmur}. 
With increase in $L_R$, the presence of the terms $(\beta_0 a_s L_R ) ^k$
makes the truncation of the perturbation series unreliable.
The solution, proposed
in this letter, resulting from RG improved resummation of  
those terms that spoil the perturbation series, shows an impressive improvement at every order. 
In fig.\ref{mur-var}, we show both the FO and the RESUM cross 
sections up to N$^3$LO for LHC13 by varying $\mu_R$ in the range $[0.1m_{\rm H}, 10m_{\rm H}]$
and keeping $\mu_F = m_{\rm H}$ fixed. For $\mu_R < m_{\rm H}$, as discussed before,
the large contributions from $L_R$ make the FO QCD corrections flip the sign
and hence the cross sections take a downturn below certain $\mu_R$. This phenomenon can 
foremost be seen for cross section at higher orders, e.g., for $\mu_F = m_{\rm H}$ the N$^3$LO cross section starts 
declining below $\mu_R = 0.5 m_{\rm H}$, followed by NNLO cross section at   
$\mu_R = 0.2 m_{\rm H}$ and so on. For $\mu_F = 2 m_{\rm H}$ also a similar pattern can be seen. For larger values 
of $\mu_R > m_{\rm H}$, however, $a_s (\mu_R^2)$ falls down suppressing the logarithmic contributions and hence the 
cross sections will decrease monotonically. 
We have also plotted the RESUM cross sections at various orders 
in Fig.~\ref{mur-var} as a function of $\mu_R$.
We find that the predictions from the RESUM cross sections
are more stable compared to the FO ones over a wide range of $\mu_R$ 
demonstrating the power and the reliability of resummation.
%
\begin{center}
\begin{table}[h!]
\begin{tabular}{| c || c | c | c | c |}
    \hline
     & ~~~LO~~~ & ~~NLO~~ & ~NNLO~ & ~N$^3$LO~ \\      
    \hline \hline
    FO (\%) & 167.26 & 143.40 & 54.99 & 27.01 \\
    \hline
    RESUM (\%) & 6.11 & 5.47 & 3.39 & 1.23 \\
    \hline
  \end{tabular}
 \caption{Percentage of maximum uncertainty for $\mu_R$ variation in the range [$0.1 m_{\rm H}, 10 m_{\rm H}$] up to N$^3$LO 
          (see text).}
 \label{table:perc}
\end{table}
\end{center}
\vspace{-0.7cm}
In Table~\ref{table:perc}, we show the maximum percentage of uncertainty in the cross sections up to 
N$^3$LO for $\mu_R$ variation in the range $[0.1 m_{\rm H}, 10 m_{\rm H}]$.
Here, at N$^3$LO, the $\mu_R$ uncertainty is maximum for $\mu_R$ between about $0.1 m_{\rm H}$ and $0.5 m_{\rm H}$
whereas at NNLO, the maximum uncertainty is for $\mu_R$ between about $0.2 m_{\rm H}$ and $10 m_{\rm H}$.
We notice that the scale uncertainties in both FO and RESUM cross sections decrease with the order of the 
perturbation theory, as expected.

\begin{figure}[h!]
\centerline{
\includegraphics[width=0.5\textwidth]{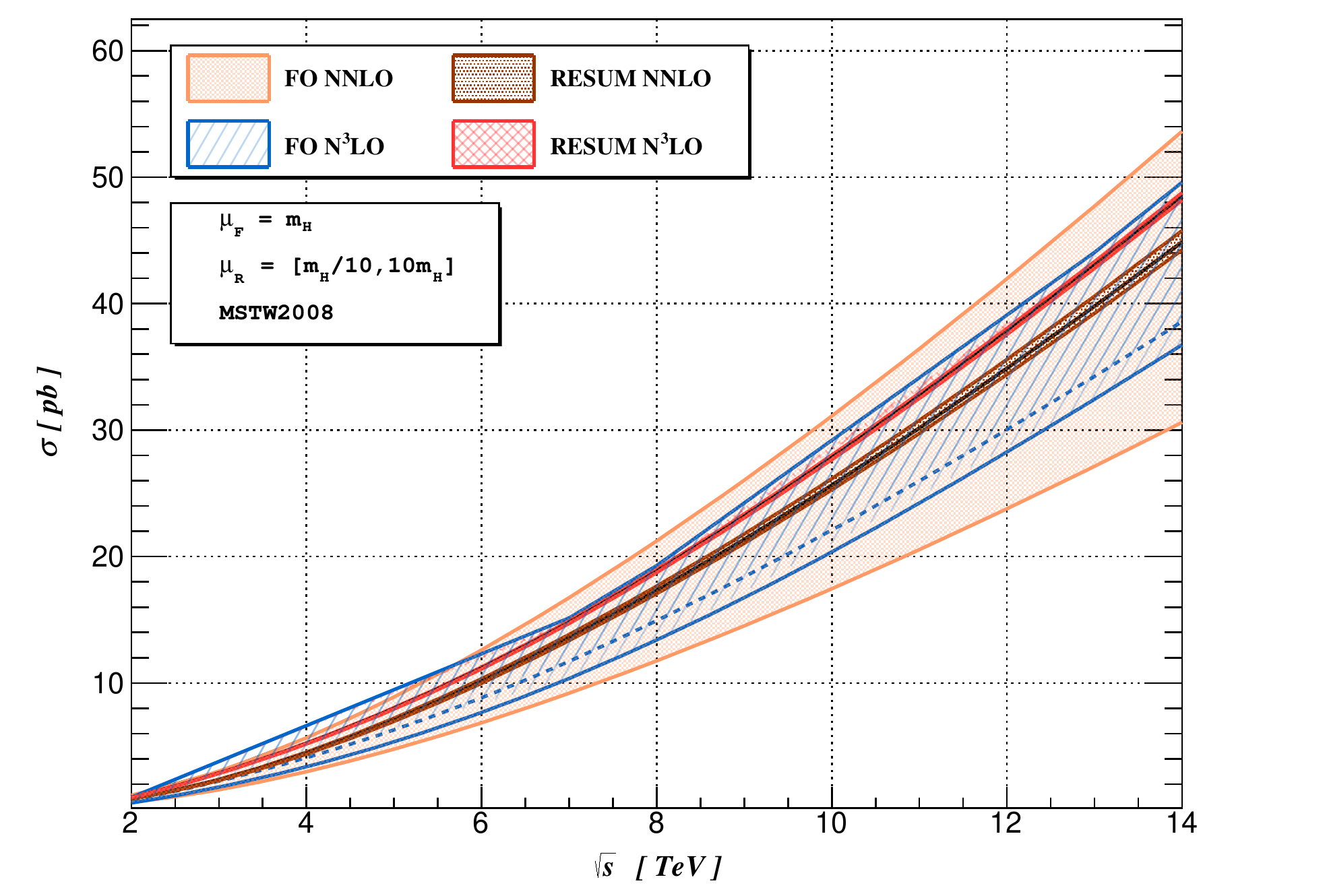}
}
\caption{\label{ecm-var} 
Dependence of scale uncertainties in both the fixed order and resummed cross sections on $\sqrt{s}$ (see text).
}
\end{figure}

We also study the scale uncertainties of both the FO and RESUM cross sections
up to N$^3$LO as a function of the center of mass energy $\sqrt{s}$ of the incoming protons 
at the LHC and our results are given in fig.\ref{ecm-var}.  Here, we vary $\mu_R$ in the 
range $[0.1m_{\rm H}, 10m_{\rm H}]$ fixing $\mu_F = m_{\rm H}$. In general, the scale uncertainties
in both FO and RESUM results are found to increase with $\sqrt{s}$ precisely because of 
the increase in gluon fluxes. Irrespective of the order of 
the perturbation theory, the RESUM results are found to decrease the scale uncertainties
remarkably compared to the FO results.  Here, at N$^3$LO, the cross sections will
increase from $\mu_R = 0.1 m_{\rm H}$ to about $\mu_R = 0.5 m_{\rm H}$ ( shown as solid lines in the 
Fig.\ref{ecm-var}, the dashed line corresponds to the one at $\mu_R = 10 m_{\rm H}$) and then start decreasing with further 
$\mu_R$ variation. Also for $\mu_R > m_{\rm H}$, the N$^3$LO cross section will decrease. 
Consequently for N$^3$LO, the cross sections at the end points of the $\mu_R$ variation i.e. $0.1m_{\rm H}$
and $10m_{\rm H}$, will both be below the one at $\mu_R = m_{\rm H}$.  

In conclusion, we have investigated the dependence of both the fixed order as well as the resummed predictions on the renormalization scale, using the recently available
results on the Higgs boson production to N$^3$LO in gluon fusion. For the resummed results,
we systematically include all the RG accessible logarithms, $L_R$, 
 to all orders in the perturbation theory.
While the fixed order N$^3$LO result shows impressive scale reduction
for the canonical choice of the renormalization scale between $m_{\rm H}/2$ and 2 $m_{\rm H}$,
there is still a significant dependence on the scale through these large logarithms which
can spoil the behavior if the renormalization scale is varied further away from this
range. On the other hand, the resummed results obtained in this letter
show little dependence on the scale choice. For $\mu_R$ in the range $[0.1 m_{\rm H}, 10 m_{\rm H}]$,
the RG improved cross sections bring the scale uncertainties from about $27$\% down to
about $1.5$\% at N$^3$LO level. This approach can also be used for other processes
such as top pair production, multi-jet production etc.\\
{\bf Acknowledgments :} We thank Nandadevi cluster computing facility at the Institute of Mathematical Sciences (IMSc) where the computation was carried out. GD thanks for the hospitality provided by IMSc
where part of the work was done. GD also thanks P. Mathews for useful discussions and for his encouragement.

\vspace{-0.5cm}


\begin{thebibliography}{99}

  
\bibitem{Anastasiou:2015ema}
  C.~Anastasiou, C.~Duhr, F.~Dulat, F.~Herzog and B.~Mistlberger,
  arXiv:1503.06056 [hep-ph].
  

\bibitem{AtlasCMS}
  G.~Aad {\it et al.}  [ATLAS Collaboration],
  Phys.\ Lett.\ B {\bf 716}, 1 (2012) ;
  S.~Chatrchyan {\it et al.}  [CMS Collaboration],
  Phys.\ Lett.\ B {\bf 716}, 30 (2012).

\bibitem{nnlo}
H.~M.~Georgi, S.~L.~Glashow, M.~E.~Machacek, and D.~V.~Nanopoulos, 
  {Phys. Rev. Lett. {\bfseries 40} (1978) 692} ;
%
A.~Djouadi, M.~Spira, and P.~M.~Zerwas, 
 {Phys. Lett. {\bfseries B 264} (1991) 440} ;
%
{S.~Dawson}, 
 {Nucl. Phys. {\bfseries B359} (1991) 283} ;
%
M.~Spira, A.~Djouadi, D.~Graudenz, and P.~M.~Zerwas, 
 {Nucl. Phys. {\bfseries B453} (1995) 17} ;
  S.~Catani, D.~de Florian and M.~Grazzini,
  JHEP {\bf 0105} (2001) 025 ;
  R.~V.~Harlander and W.~B.~Kilgore,
  Phys.\ Rev.\ D {\bf 64} (2001) 013015 ,
%
  {Phys. Rev. Lett. {\bfseries 88} (2002) 201801} ;
%
C.~Anastasiou and K.~Melnikov, 
  {Nucl. Phys. {\bfseries B646} (2002) 220} ;
%
{V.~Ravindran, J.~Smith, and W.~L.~van Neerven}, 
  {Nucl. Phys. {\bfseries
  B665} (2003) 325}.

  
  
\bibitem{nnll} 
S.~Catani, D.~de~Florian, M.~Grazzini, and P.~Nason, 
 {JHEP {\bfseries 0307} (2003) 028}.  
  

\bibitem{ewnnlo}
  U.~Aglietti, R.~Bonciani, G.~Degrassi and A.~Vicini,
  Phys.\ Lett.\ B {\bf 595} (2004) 432 ;
  S.~Actis, G.~Passarino, C.~Sturm and S.~Uccirati,
  Phys.\ Lett.\ B {\bf 670} (2008) 12 .

  

%

\bibitem{Anastasiou:2014vaa}
  C.~Anastasiou, C.~Duhr, F.~Dulat, E.~Furlan, T.~Gehrmann, F.~Herzog and B.~Mistlberger,
  Phys.\ Lett.\ B {\bf 737} (2014) 325 .
  

\bibitem{oursn3loSV}  
  T.~Ahmed, M.~Mahakhud, N.~Rana and V.~Ravindran,
  Phys.\ Rev.\ Lett.\  {\bf 113} (2014) 11,  112002 ;
%
  T.~Ahmed, M.~K.~Mandal, N.~Rana and V.~Ravindran,
  Phys.\ Rev.\ Lett.\  {\bf 113} (2014) 212003 ;
%
%
  S.~Catani, L.~Cieri, D.~de Florian, G.~Ferrera and M.~Grazzini,
  Nucl.\ Phys.\ B {\bf 888} (2014) 75 ;
%
  T.~Ahmed, N.~Rana and V.~Ravindran,
  JHEP {\bf 1410} (2014) 139 ;
%
  T.~Ahmed, M.~K.~Mandal, N.~Rana and V.~Ravindran,
  JHEP {\bf 1502} (2015) 131 ;
%
  M.~C.~Kumar, M.~K.~Mandal and V.~Ravindran,
  JHEP {\bf 1503} (2015) 037.
%


   
  
  
\bibitem{Baglio:2010um}
  J.~Baglio and A.~Djouadi,
  JHEP {\bf 1010} (2010) 064 .

  
  
\bibitem{Ahrens:2008qu}
 V.~Ahrens, T.~Becher, M.~Neubert and L.~L.~Yang,
 Phys.\ Rev.\ D {\bf 79} (2009) 033013 .
  
  
\bibitem{Ahrens:2008nc}
  V.~Ahrens, T.~Becher, M.~Neubert and L.~L.~Yang,
  Eur.\ Phys.\ J.\ C {\bf 62} (2009) 333 .
  
  
\bibitem{Ahmady:2002fd}
  M.~R.~Ahmady, F.~A.~Chishtie, V.~Elias, A.~H.~Fariborz, N.~Fattahi, D.~G.~C.~McKeon, T.~N.~Sherry and T.~G.~Steele,
  Phys.\ Rev.\ D {\bf 66} (2002) 014010 .
  
  
\bibitem{Moch:2005ba}
  S.~Moch, J.~A.~M.~Vermaseren and A.~Vogt,
  Nucl.\ Phys.\ B {\bf 726} (2005) 317 .
  
  
\bibitem{Martin:2009iq}
  A.~D.~Martin, W.~J.~Stirling, R.~S.~Thorne and G.~Watt,
  Eur.\ Phys.\ J.\ C {\bf 63} (2009) 189 .
  
  
\bibitem{Whalley:2005nh}
  M.~R.~Whalley, D.~Bourilkov and R.~C.~Group,
  hep-ph/0508110.


\end{thebibliography}
\end{document}